\documentclass[aps,prl,twocolumn,showpacs]{revtex4-1}

\usepackage{amsmath}
\usepackage{bm}
\usepackage{graphicx}

\def\ii{\textrm{i}\,}

\begin{document}

\title{Spin selective transport through helical molecular systems}

%

\author{R.\ Gutierrez$^{1}$}
\author{E.\ D\'{\i}az$^{1,2}$}
\author{R.\ Naaman$^{3}$}
\author{G.\ Cuniberti$^{1,4}$}
\affiliation{$^{1}$Institute for Materials Science, Dresden University of Technology, 01062 Dresden, Germany\\
$^{2}$GISC, Departamento de F\'{\i}sica de Materiales, Universidad Complutense, E-28040 Madrid, Spain\\
$^{3}$Department of Chemical Physics, Weizmann Institute, 76100 Rehovot, Israel\\
$^{4}$Division of IT Convergence Engineering National Center for Nanomaterials Technology, POSTECH, Pohang 790-784, Republic of Korea}

\begin{abstract}
Highly spin selective transport of electrons through a helically shaped electrostatic 
potential is demonstrated in the frame of a minimal model approach. The effect is significant 
even for weak spin-orbit coupling.  Two main factors determine the selectivity, 
an unconventional Rashba-like spin-orbit interaction, reflecting the helical symmetry 
of the system, and a weakly dispersive electronic band of the helical system. The weak  
electronic coupling, associated with the small dispersion,  leads to a low mobility of the charges 
in the system and allows even weak spin-orbit interactions to be effective. 
The results are expected to be generic for chiral molecular systems 
displaying low spin-orbit coupling and low conductivity.
\end{abstract}

\pacs{
73.22.-f 
73.63.-b 
72.25.-b 
87.14.gk 
87.15.Pc 
}
\maketitle

{\textit{Introduction$-$}}The concept of spintronic devices operating without a magnetic 
field has been proposed some time ago for solid state devices in which the spin-orbit 
coupling (SOC) is large~\cite{datta90,lu98}. Recently,  a new type of magnet-less spin selective 
transmission effect has been reported~\cite{ray99,naaman10,ray06,goehler11}.
It was found  that electron transmission through chiral molecules is highly spin selective at room temperature.
These findings are surprising since carbon-based molecules have typically a 
small SOC that cannot support significant splitting between the spin states, 
splitting which is thought to be essential for any spin dependent property. 
Although it has been found both in theory~\cite{ando08,huertas06,demartino02} and 
experiments~\cite{kuemmeth08} that there 
is a cooperative contribution to the value of the SOC, so that this quantity 
may be larger in molecules or nanotubes than in a single carbon atom, the values 
calculated or experimentally found are still relatively 
small~\cite{ando08,huertas06,demartino02,kuemmeth08,rashba}, e.g.  few meV for carbon nanotubes
~\cite{kuemmeth08}. Hence, even including this cooperative contribution, 
the values obtained for the spin polarization (SP) in electron transmission through 
chiral molecules~\cite{goehler11} seem to be too high and cannot be rationalized by such SOC values. 

Recently, a theoretical model based on the first Born approximation in scattering theory has been proposed for explaining the spin selectivity of chiral molecules~\cite{yeganeh09}. 
Although the results are in qualitative agreement with the experimental observations, they could not explain them using reasonable SOC values. 

In what follows, a model is presented to describe electron transmission through a 
helical electrostatic potential (see Fig.~1). Although the model does 
not claim to fully catch the complexity of the experimentally studied DNA-based systems~\cite{goehler11,xie},
it highlights the role of some crucial parameters, which can determine the experimentally observed high SP. 
The key factors in the model that allow for the high spin selectivity are: 
i) Lack of parity symmetry due to the chiral symmetry of the scattering potential; 
ii) Narrow electronic band widths in the helical system, i.e. the interaction 
between the units composing the helical structure through which the electron 
propagates is relatively weak. Moreover, a physically meaningful estimation of the SOC can be obtained by taking into account that first, like in the solid state, in the 
present study the electric field acting on the electron needs to include the 
effective influence of all the electrons belonging to a molecular unit~\cite{xie,bihlmayer07}, and second, 
due to proximity effects, the Coulomb 
interaction between the transmitted electron and the atoms in the molecular unit can 
scale as $1/R$ for short distances $R$. 

\begin{figure}[b]
\centerline{\includegraphics[height=0.25\textwidth,clip]{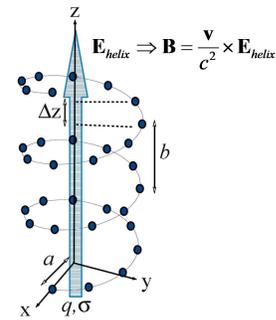}}
\caption{A charge $q$ in spin state $\sigma$ 
is moving along through helical electric field. 
The parameters $a$, $b$ and $\Delta z$ are the radius and the pitch of the helix and 
the spacing of the z-component of the position vector of the charges 
distributed along it, respectively. The helical field ${\bf E}_{\rm helix}$ 
induces a magnetic field ${\bf B}$ in the rest frame of the charge and 
hence influences its spin state.} 
\label{helix}
\end{figure}
{\textit{Model and Methodology$-$}}We consider the Schr\"{o}dinger equation for a particle 
moving in a helical electrostatic field. Analytical results for such fields have been 
derived in Ref.~\onlinecite{hochberg97}. For the sake of simplicity, 
approximate expressions valid near 
the z-axis will be used (only $x$ and $y$ components will be considered, the $z$ component only contributes when considering the full three-dimensional problem, see below ): 
${\bf{E}}_{\textrm {helix}} =-E_0 \sum_{i,j} g_{i,j}(z)(\cos(Qj\Delta z),\sin(Qj\Delta z))$.
Here, $g_{i,j}(z)=(1+[(z-ib-j\Delta z)/a]^2)^{-3/2}$ and $Q=2\pi/b$ 
with $b$ being the helix pitch and $a$ the helix radius, see Fig.~\ref{helix}. 
The index $m=0,\cdots,M_{0}-1$ runs along one helical turn and labels the z-coordinate of the $M_{0}$ molecular units
 placed along the helix. The index $n=-L_{0}/2,\cdots,L_{0}/2$ ($L_{0}$ being the number of helical turns) connects sites which differ in their z-coordinate by $b$~\cite{footnote}. We note that the considered helical potential is assumed to be related to the charge distribution along the stack of molecular units building the helical structure; hence the factor $E_0$ is proportional to the local charge density.

\begin{figure}[t]
\centerline{\includegraphics[width=0.33\textwidth,clip]{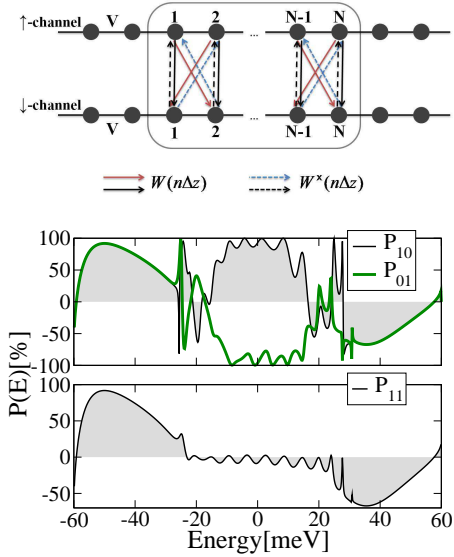}}
 \ \\
\centerline{\includegraphics[width=0.33\textwidth,clip]{Figure_2b.eps}}
\caption{\textit {Top panel}: Schematic representation of the tight-binding model, see Eq.~\ref{TB}. 
The two channels interact via the SOC (framed region). To the left and right of the spin scattering region, 
both channels are independent and are modeled by semi-infinite chains. \textit {Bottom panel}: 
Energy dependence of the SP $P(E)$ for $L_{0}$=3  helical turns, 
and for injected electrons  polarized with their spin pointing up ($P_{10}$), 
down ($P_{01}$), or unpolarized  ($P_{11}$). 
A spin-filter effect takes place only for energies near the band edges, where all SPs
have the same sign. Notice also that near the 
band edges the SP has opposite signs for electrons ($E<0$) and holes ($E>0$), 
though $P(E)$ is not exactly antisymmetric. Parameters: 
$\alpha=5$ meV nm, $V=30$ meV, $U_{0}= 3$ meV. 
}
\label{ladder} 
\end{figure}

For a charge moving with momentum ${\bf p}$ through the helix, 
the field ${\bf E_{\rm helix}}$ induces a magnetic field in the charge's rest frame, 
from which a SOC arises: 
$H_{SO}=\lambda \pmb {\sigma}({\bf p}\times {\bf E}_{\rm helix})$.
The SOC strength is  $\lambda = e \hbar / 4m^2 c^2$ and $\pmb {\sigma}\ $ is a vector 
whose components are the Pauli matrices $\sigma_x$, $\sigma_y$, $\sigma_z$. 
The  general problem is three-dimensional; however, in order to get first insights into 
the behavior of the SP, we will assume $p_{x}=p_{y}=0, p_{z}\neq 0$, so that the Schr\"odinger equation takes the form~\cite{si}:
\begin{eqnarray}
\Big[&-&\frac{\hbar^2}{2m}\partial_z^2+ U(z)+\alpha \begin{pmatrix} 0&\Psi(z) \\ -\Psi^*(z)&0 \end{pmatrix}\partial_z\nonumber\\
&-&\alpha \begin{pmatrix} 0&f(z)\\ f^*(z)&0 \end{pmatrix} \Big] {\bf {\chi}}(z)=E{\bf {\chi}}(z).
\label{hamiltonian}
\end{eqnarray}
Here, ${\bf {\chi}}(z)=( {\chi}^{\uparrow}(z), {\chi}^{\downarrow}(z))^{T}$ is a spinor, $\Psi(z)=E_{x}-\ii E_{y}=\sum_{i,j} e^{-\ii Qj\Delta z}g_{i,j}(z)$, $f(z)=\partial_z\Psi(z)$, and $U(z)$ the helical electrostatic potential. The terms $\sim f(z),f^{*}(z)$ are introduced to make the Hamiltonian hermitian in the continuum representation. The SOC parameter $\alpha=\hbar \lambda E_0$ (with dimensions of energy$\times$length) depends on the effective charge density through $E_0$. The problem posed by Eq.~\ref{hamiltonian} can be written as an effective two-channel 
nearest-neighbor tight-binding model~\cite{si}:
\begin{eqnarray}
H&=&\sum_{\sigma=\uparrow,\downarrow}\sum_{n=1}^{N} U_{n} c_{n,\sigma}^{\dagger}c_{n,\sigma}
+ V\sum_{\sigma=\uparrow,\downarrow}\sum_{n=1}^{N-1}(c_{n,\sigma}^{\dagger}c_{n+1,\sigma}+{\rm h.c.})\nonumber\\
&+&\sum_{n,m=1}^N  (c_{n,\uparrow}^+W_{n,m}c_{m,\downarrow}+c_{m,\downarrow}^{+}W_{m,n}^{\times}c_{n,\uparrow})+H_{leads}. 
\label{TB}
\end{eqnarray}
The operators $\{c_{n,\sigma},c_{n,\sigma}^+\}_{n=1,\dots,N,\sigma=\uparrow,\downarrow }$ create or destroy, respectively, an 
excitation at the tight-binding site $n$ with spin index $\sigma$.  The only non-zero elements of the  inter-channel coupling matrix ${\bf W}$ 
are given by~\cite{si}:
$W_{n,n}=-\alpha f(n \Delta z)$, $W_{n,n+1}=\alpha \Psi(n\Delta z)/2\Delta z$, and $W_{n+1,n}=-\alpha \Psi((n+1)\Delta z)/2\Delta z$. Further, the matrix  $W_{n,m}^{\times}$ satisfies $W_{n,m}^{\times}=-(W_{n,m})^*$ for $n\neq m$, and $W_{n,n}^{\times}=(W_{n,n})^*$. 
The hopping  $V$ should in general be estimated on the basis of a 
first-principle calculation of the electronic coupling for a given system. 
However,  we will consider it as a free parameter, whose order of magnitude 
for helical organic systems is expected to lie in the range of few tens of meV (e.g. for DNA, electronic structure calculations yield values of the 
order of $20-40$ meV~\cite{woic09}). Finally, the operator $H_{leads}$ includes the semi-infinite chains  to the left (L)  and right (R) of the SO active region~\cite{si}. A schematic representation of this  two-channel 
model is shown on the top panel of Fig.~\ref{ladder}.

{\textit{Transport properties$-$}}We focus on the spin-dependent transmission probability, $T(E)$, of the model Hamiltonian given by Eq.~\ref{TB}, as a function of the electron's injection energy $E$. 
The problem can be considered as a scattering problem where a finite-size region (with non-vanishing  SOC) is  coupled to two independent  $L$ (left)- and two independent $R$(right)-electrodes, each electrode standing for a spin 
channel and being represented by a semi-infinite  chain, see Fig.~\ref{ladder}. $T(E)$ encodes the influence of multiple scattering events in the SOC region; using Landauer's theory~\cite{nitzan01} we obtain~\cite{si}:
\begin{eqnarray}
T(E)&=&\Gamma_{\uparrow}^R(\Gamma_{\uparrow}^L|G_{1\uparrow,N\uparrow}|^2+\Gamma_{\downarrow}^L|G_{1\downarrow,N\uparrow}|^2)\nonumber \\
&+&\Gamma_{\downarrow}^R(\Gamma_{\uparrow}^L|G_{1\uparrow,N\downarrow}|^2+\Gamma_{\downarrow}^L|G_{1\downarrow,N\downarrow}|^2)\nonumber \\
&=&t_{up}(E)+t_{down}(E)\ .
\label{transmision}
\end{eqnarray}
In Eq.~\ref{transmision}, $G_{n\sigma,m\nu}(E)$ with $\sigma,\nu=\uparrow,\downarrow$ are matrix elements of the retarded Green's function of 
the SOC region including the influence of the $L$- and 
$R$-electrodes. The individual contributions in Eq.~\ref{transmision}
can be related to different transport processes without (e.g. $\Gamma_\uparrow^L\Gamma_\uparrow^R|G_{1\uparrow,N\uparrow}|^2$) or 
with (e.g. $\Gamma_\uparrow^L\Gamma_\downarrow^R|G_{1\uparrow,N\downarrow}|^2$) spin-flip scattering, see  Fig.~\ref{ladder}. 
 Notice that $t_{up}(E)$ and $t_{down}(E)$ 
$-$the transmissions for the up and down channels respectively, as defined by 
Eq.~\ref{transmision}$-$, contain  contributions arising both from direct transmission without 
spin-flip as well as spin-flip. 
An energy-resolved SP for 
different initial spinor states can be defined as: $P(E)=(t_{up}(E)-t_{down}(E))/T(E)$.
The energy-average SP
$\langle P(E)\rangle_E=P(\langle t_{up}(E)\rangle,\langle t_{down}(E)\rangle,\langle T(E)\rangle)$ 
will also be used. We focus only on electron-like contributions ($E<0$) 
and on energies $|E|\ge k_BT\approx23$ meV, so that  $\langle\dots\rangle_E=\int_{-2V}^{-k_BT}dE(\dots)$.

{\textit{Results$-$}}A crucial parameter in the model is the SOC coupling  $\alpha$. Realistic values are obviously very difficult to obtain~\cite{soc1,soc2}, since $\alpha$ 
is not simply the  atomic SOC, but contains the 
influence of the charge distribution in the system via the field factor $E_0$. 
For the sake of reference, a  rough value of $E_0$ for DNA may be estimated along the following lines. 
A single DNA base is considered as composed of discrete point-like charge centers $A$, 
representing the atoms. We associate with each center $A$ at position $R_A$ a 
Gaussian-shaped charge distribution of width $w\sim0.3-0.4$ nm and with strength given by an estimated atomic charge 
density $\rho_0$ for C, N, and O atoms (considered as spheres with 
a radius of the order of the corresponding covalent radius). The local  
field of this charge distribution, $E_0=-(1/4\pi\epsilon_0)(\partial/\partial r)\int d^3r' \rho(r'-R_A)|r-r'|^{-1}$, 
can be computed analytically~\cite{si} and it scales for  $R=|r-R_A|\ll w$ like 
$E_0\approx (N_{0}\rho_{0}/4\pi\epsilon_0)(w/2\sqrt \pi)^2 R^{-1}$ ($E_0$ has been multiplied by a factor $N_{0}\sim 10$, the number of atoms in a base, to approximately  account for other charge centers. 
For $R/w\sim 0.3-0.4$, values of $\alpha=\hbar \lambda E_0\approx 1.87-2.35$ meV nm can be obtained. In the calculations, $\alpha\sim 2-6$ meV nm have been  used. 
Though the previous discussion provides a very rough estimate, it highlights the need of considering the influence 
of many charges through  $\rho_{0}$ and $N_{0}$ as well as proximity effects (short-distance scaling of $E_0$) in the estimation of $\alpha$.

Fig.~\ref{ladder} presents the energy dependent SP for 
different incoming spin states when the spin is pointing up (10), down (01) or the electrons 
are unpolarized (11).
The coupling $\alpha$
was assumed to be $5$ meV nm. Although this value is larger than the previously estimated 
one, it serves to illustrate the behavior of the model in a clear way. 
In the case of (10) and (01) states, the interesting energy 
windows are those where both SPs have the same sign, which indicates that 
the outgoing state will  always have the same SP  \textit{independently} of the 
initial condition. This behavior 
occurs mainly for energies near the band edges. A similar situation is found for the (11) state, see Fig.~\ref{ladder}.
 Near the band center, $P_{10}(E)$ and $P_{01}(E)$ have opposite signs and 
hence the SP depends on the incoming spin state. 
The average SPs, as defined above,  amount  to approximately $\langle P_{10}\rangle_{E}=\langle P_{01}\rangle_{E}= \langle P_{11}\rangle_{E}\approx$62\%. 

\begin{figure}[t]
\centerline{\includegraphics[width=0.45\textwidth]{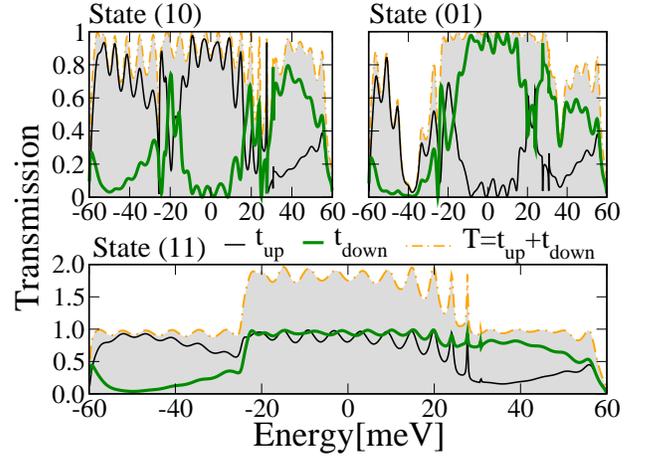}}
\caption{Different components of the transmission  
$t_{up}(E)$,$t_{down}(E)$ and $T(E)$ as defined in Eq.~\ref{transmision}, and for the same parameters of Fig.~\ref{ladder}. 
Focusing on electron-like contributions, it is only near the lower band edge ($E\leq -22$ meV) where 
a positive  SP for all incoming states (10), (01), and (11) is obtained, see also Fig.~\ref{ladder}.
}
\label{T} 
\end{figure}

Figure~\ref{T} shows the corresponding spin-resolved transmissions. Notice first, that the states  (10) and (01) 
correspond to cases where one of the incoming spin channels is decoupled from the 
system by setting the corresponding $\Gamma^L=0$~\cite{si},
and hence, the total transmission cannot exceed one.  
For (11) both channels are open and the maximum transmission is $2$. 

In the top panel of Fig.~\ref{T}, for (10) and (01), we find some degree of spin-dependent 
back-scattering, which is reflected in the different total transmissions $T(E)$ 
for each polarization.
In what follows, for the sake of reference, only the behavior in the energy window $[-2V,-k_{B}T], k_{B}T\sim 23$ meV is discussed. 
For the (10) state, transmission without spin flip is dominant in this energy region, and this leads to the positive SP. However, for (01), spin-flip processes become dominant in the same energy region, and hence the outgoing up-channel acquires a larger weight. As a result the SP for (01) is also positive. This behavior is closely related to the chiral symmetry, which basically manifests in the special structure of the $W,W^{\times}$ matrices. For the (11) state, bottom panel of Fig.~\ref{T}, the outgoing up-channel clearly dominates the transmission in the considered energy window , thus indicating that for unpolarized electrons back-scattering and spin-flip of the down-component will ultimately lead  to a positive SP. A similar analysis can be performed for the hole-like energy region $E>0$.
In general terms, SP may occur either by spin-flip  (with  no net change of the total transmission) or by spin selective back-scattering. The results 
of Fig.~\ref{T} suggest that both processes are playing a 
role; their relative contribution to the SP turns out  however to sensitively depend on the 
specific energy window considered. 
\begin{figure}[t]
 \centerline{\includegraphics[width=0.40\textwidth,clip]{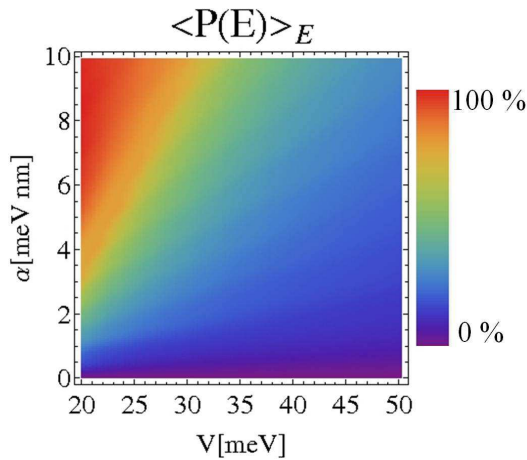}}
\caption{2D plot of the energy average SP  $\langle P(E)\rangle_E$ as a function of both
 the hopping parameter $V$  and the SOC $\alpha$. Only for small $V$ a relative large SP is found. With increasing electronic coupling, larger SOC strengths are required to get a sizeable SP.
}
\label{2D} 
\end{figure}
The selectivity found in this model  relates to two special features of the chiral system: 
(i) the  symmetry of the field  which translates into an unconventional SOC, and 
(ii) the narrow electronic band width in chiral organic systems. 
The term band width serves only as a keyword for the averaged value of the  coupling matrix elements, $V$, 
between neighboring molecular states mediating 
charge motion.
As shown in Fig.~\ref{2D}, the size of the hopping parameter 
strongly affects the energy average SP, ultimately leading to $\langle P(E)\rangle_E\rightarrow0$ for large $V$. For small hopping, however, the SP can achieve very large values by only a moderate increase of the SOC $\alpha$. 
The interplay between $\alpha$ and $V$ seems  related 
to the relatively long time (roughly proportional to $\hbar V^{-1}$) 
the electron  will spend in the conducting channel in a real system, allowing for the SOC 
to become more effective. 

{\textit{Conclusions$-$}}The present study based on a generic model 
sheds new light on a chirality-induced spin selectivity (CISS) effect. It suggests that 
beyond the symmetry itself, CISS depends on the organic molecules being poor conductors. 
Weak electronic coupling along the helical structure 
is expected to lead to low mobility of the electrons through the system and allows enough time for the SOC, 
although being weak, to influence spin transport. The effect depends on the 
electron momentum and once the electrons have kinetic energy above $k_BT$, the SP 
increases and becomes weakly energy dependent. One open issue for 
further inquiry is the influence of the electrode-molecule interface. If the electrodes are magnetic, spin-dependent tunnel barriers emerge, which may
influence the SP. The present study indicates that CISS may 
be a very general phenomenon, existing in chiral systems having low SOC and low conductivity, 
and hence may play a role in charge transport through biosystems. The effect could also be of great interest to control the spin injection efficiency in the context of semiconductor-based spintronics by interfacing chiral molecules with semiconductor materials.

\begin{acknowledgments}
RG and ED thank H. Pastawski, R. Bustos-Marun, T. Brumme, and S. Avdoshenko 
for fruitful discussions. This work was partially funded by the DFG under CU 44/20-1, MAT2010-17180 and by the South Korea Ministry of 
Education, Science, and Technology Program ``World Class University'' (No. R31-2008-000-10100-0). 
Computational resources were provided by the  ZIH at TU-Dresden. ED thanks MEC and RN thanks the German-Israel Science Foundation and  the Israel Science Foundation for financial support. 

\end{acknowledgments}


\begin{thebibliography}{99}
\bibitem{footnote} We chose for the sake of reference $b\sim3.2$ nm, $a\sim0.7$ nm, and $\Delta z=0.32$ nm, values which are  close to those of B-DNA.
\end{thebibliography}
\end{document}